\documentstyle[epsf]{article} 
\newcommand {\etal} {{\it et al. }}
\begin{document}

{\LARGE A confirmed location in the Galactic halo for the high-velocity
cloud 'chain A'}


Hugo van Woerden$^a$, Ulrich J. Schwarz$^a$, Reynier F. Peletier$^{a.b}$, 
Bart P. Wakker$^c$, Peter M.W. Kalberla$^d$
\bigskip

a Kapteyn Institute, Postbus 800, 9700 AV Groningen, The Netherlands 
\hfil\break
b Dept. of Physics, University of Durham, South Road, Durham DH1 3LE, UK
\hfil\break
c Department of Astronomy, University of Wisconsin, Madison WI 53706, USA 
\hfil\break
d Radio-astronomisches Institut, Universit\"at Bonn, 53121 Bonn, Germany 
\hfil\break
\bigskip


The high-velocity clouds of atomic hydrogen, discovered about 35 years 
ago$^{1,2}$, have velocities inconsistent with simple Galactic rotation 
models that generally fit the stars and gas in the Milky Way disk.  Their 
origins and role in Galactic evolution remain poorly understood$^3$,
largely for lack of information on their distances.  The high-velocity 
clouds might result from gas blown from the Milky Way disk into the halo 
by supernovae$^{4,5}$, in which case they would enrich the Galaxy with 
heavy elements as they fall back onto the disk.  Alternatively, they may 
consist of metal-poor gas -- remnants of the era of galaxy 
formation$^{2,6-8}$, accreted by the Galaxy and reducing its metal 
abundance.  Or they might be truly extragalactic objects in the Local 
Group of galaxies$^{7-9}$.  Here we report a firm distance bracket for 
a large high-velocity cloud, Chain A, which places it in the Milky Way 
halo (2.5 to 7 kiloparsecs above the Galactic plane), rather than at an 
extragalactic distance, and constrains its gas mass to between 10$^5$ 
and 2 $\times$ 10$^6$ solar masses.

    
Distance estimates of HVCs have long been based on models or indirect 
arguments$^{2,10}$. 
The only direct method uses the presence or absence of interstellar 
absorption lines at the HVC's velocity in spectra of stars at different 
distances.  Presence of absorption shows the HVC to lie in front of the 
star; absence places it beyond, provided the expected absorption is well 
above the detection limit$^{11}$.  Blue stars are best, since their spectra
contain few confusing stellar lines.  The method requires that metal ions 
are present in HVCs.  Indeed, since suitable spectrographs 
have become available, CaII and other metal-ion absorption lines have been 
found for many HVCs$^3$ in the spectra of background quasars or Seyfert 
galaxies.   Using HI column densities measured at high angular resolution, 
the metal-ion/HI ratios thus derived provide estimates of expected 
absorption-line strengths towards stars probing the same HVC. 

The HVCs MII-MIII, for which an upper limit to the distance, $d$ $<$ 4 kpc, 
is known$^{12,13}$,
but no lower limit$^{13}$, were thus found to be Galactic, 
and may even lie in the Disk, as does the tiny object$^{14,15}$ 
HVC 100-7+100, with $d$ $<$ 1.2 kpc, and  distance from the plane 
$|z|$ $<$ 0.14 kpc.  Lower distance limits are known for Complex C$^{16,17}$ 
($d$ $>$ 2.5 kpc, and probably$^{18}$ $d >$ 5 kpc),
Cloud 211 = HVC267+20+215$^{16}$ ($d >$ 6 kpc),
parts of the AntiCenter Complexes$^{19}$ ($d$ $>$ 0.6 kpc),
and Complex H$^{20}$ ($d$ $>$ 5 kpc). 
Chain A is the first HVC for which both a significant upper and a non-zero 
lower limit to its distance are known, constraining its location relative 
to the Galaxy's major components.


HVC-Complex A, also called "Chain A", was the first HVC discovered
and has been studied in detail$^{3,21}$.  It is a 30$^{\rm o}$ long 
filament, containing several well-aligned concentrations with velocities 
(relative to the local standard of rest, LSR) between -- 210 and 
-- 140 km/s.  HST spectra$^{22}$ 
of the Seyfert galaxy Mark 106 show strong MgII absorption by this HVC.
The fact that such absorption is not detected (it is less than 
0.03 times the expected strength) in the star PG0859+593, although 
the HVC's H{\sc I} emission (as measured at 1 arcmin resolution) is similar
in both directions, sets a firm lower distance limit, $d > 4 \pm 1$ kpc 
($z > 2.5 \pm 0.6$ kpc), for Chain A.  

We have now also measured an upper limit, $d$ $<$ 10 kpc, for the upper end 
of Chain A, using the RR Lyrae star AD Ursae Maioris, which lies at a 
distance of $10.1 \pm 0.9$ kpc (Fig. 1).  Detection of interstellar lines 
in the spectra of RR Lyr stars is generally hampered by the presence of many 
stellar lines; but these are fewer during maximum phase, when the star is 
hotter.  Figure 2 shows portions of the spectrum of AD UMa around the 
Ca{\sc II}-H and K lines, observed during maximum, and a 21-cm profile 
taken in the same direction.  The latter has  components at velocities $v$ 
(relative to the LSR) of --4, --40 and --158 km/s.  The Ca{\sc II}-K line 
shows the same interstellar components as the 21-cm profile, plus a strong 
stellar absorption around +70 km/s.  The weaker Ca{\sc II}-H line, though 
blended with a broad stellar H-$\epsilon$ absorption, shows similar 
structure.  In particular, absorption by the HVC at 
$v$ $\sim$ --160 km/s is  present at both K and H.


Could these HVC absorptions be affected by blending with stellar lines?  
Figure 3 shows three Fe{\sc I} lines of multiplet number 4, indicating a 
stellar radial velocity $v$ =  + 77 $\pm$ 2 km/s.  Comparison of their
strengths and widths with those$^{23}$ in the blue field-horizontal-branch 
star HD 161817 allows calculation (Fig. 4) of the profile of a fourth line 
(laboratory wavelength $\lambda_0$ = 3930.3 $\rm \AA$), predicted to be 
present at 3931.3 $\rm \AA$.  Subtraction of this predicted line from the 
observed spectrum (Fig. 4) leaves a narrow absorption at 
$v$ = -- 158.2 $\pm$ 1.2 km/s, in close 
agreement with the HVC velocity $v$ = -- 157.6 $\pm$ 0.2 km/s, found at 
21 cm (Fig. 2).  If this absorption were due to the Fe{\sc I} 
3930.3 $\rm \AA$ line, it would have $v$ = + 96 $\pm$ 1 km/s in that frame, 
which is clearly incompatible with the stellar velocity of + 77 $\pm$ 2 km/s 
found from the other lines in the same multiplet.  The velocity difference  
of 19 $\pm$ 3 km/s, the lack of any other suitable identification, 
and the agreement with the 21-cm velocity, convincingly 
show that the deep line at 3931.5 $\rm \AA$ in Figures 3 and 4 must be due 
to Ca{\sc II}-K absorption at --158 km/s by the HVC, while the shortward 
wing is due to the stellar Fe{\sc I} line.

The agreement in velocity of the high-velocity absorptions at K and H, and 
the fact that the ratio of line depths is about 2 : 1, as expected, further 
strengthens the identification of the HVC absorption.  Thus, it is certain  
that Chain A lies in front of AD UMa.  The CaII and HI line strengths 
indicate a Ca$^+$ abundance of order 0.01 times the total solar Ca 
abundance, confirming our earlier$^{11}$ tentative result.  (Note that 
interstellar calcium is generally strongly depleted by inclusion into dust 
grains, and Ca$^+$ is not the dominant ion in the interstellar gas phase.)


The absorption seen in AD UMa sets an upper limit of 10 $\pm$ 1 kpc to the 
distance of Chain A.  Combining this with the lower limit$^{22}$ of 
4 $\pm$ 1 kpc,  we conclude that the high-latitude end of Chain A lies at 
4 $<$ $d$ $<$ 10 kpc, or 2.5 $<$ $z$ $<$ 7 kpc above the Galactic plane.  
Using the H{\sc I} flux$^{21}$, we derive an H{\sc I} mass of 
9800 $d^2$ M$_\odot$, i.e. between 1.5 and 10 $\times$ 10$^5$ M$_\odot$.  
With the standard helium abundance, the (H{\sc I} + He) gas mass would be 
a factor 1.4 higher.  The nondetection$^{24}$ of CO emission from bright 
HI cores in Chain A implies that the H$_2$ mass must be one or more orders 
of magnitude less.  Recent H$\alpha$ observations$^{25}$ 
suggest that Chain A is mostly neutral.  Assuming the ionized contribution 
to be minor, the total gas mass in Chain A lies between 2 $\times$ 10$^5$ 
and 2 $\times$ 10$^6$ M$_\odot$.  The kinetic energy of the complex then is 
of order (0.3 - 3) $\times$ 10$^{53}$ erg, if we assume$^{21}$ a peculiar 
velocity, $v_{\rm {dev}}$, of -- 130 km/s.


Our distance bracket places Chain A definitely in the Galactic Halo, rather 
than in intergalactic space.  It excludes models for its nature and origin 
requiring a distance of order 1 kpc or less, such as relationships to local 
molecular clouds$^{26}$,
or collision of an intergalactic cloud with the Galactic Disk$^{27}$. 
It also rules out that Chain A would be a 
Galactic satellite at about 50 kpc distance$^{9}$, 
or a protogalactic gas cloud at $\sim$ 500 kpc distance$^{28}$, 
or "a member of the Local Group of galaxies", 
as proposed recently$^{7}$ for HVCs in general. 
Other HVCs may well be at such great distances and fit the latter model; 
and some, e.g. the tiny, nearby cloud HVC100-7+100 (see above), may have 
a local origin$^{15}$.

The location of Chain A in the Halo still allows several models for its 
origin.  For its height $2.5<z<7$ kpc to be consistent with a 
Galactic-Fountain model$^{4,5}$, 
a sufficiently hot halo ($T > 5 \times 10^5$ K) would be required.  The 
small-scale structure observed$^{29,30}$ in Chain A 
would then be due to instabilities formed in the downward flow of cooling 
clouds.  Alternatively, Chain A may represent gas captured from intergalactic 
space$^{2,6,8}$. 
In that case, collision with an ionized halo extending to high $z$ may have 
served to decelerate the gas to its present velocity$^{2,6,31}$, and to form 
the small-scale structure, which has typical time-scales$^{3,29}$ 
of order 10$^7$ years, and therefore probably formed within a 
few kpc of its present location.  In this accretion model, the question 
whether the origin of Chain A lies in the Magellanic System (as debris from 
encounters between Milky Way and Magellanic Clouds), or far away in the Local 
Group (as "remnant of Local Group formation"$^{7,8}$), 
remains open: location in the Galactic Halo does not preclude such a 
distant origin.

A clue to the origin of Chain A might be found in its metallicity.  
In a Galactic Fountain, near-solar metallicities would be expected;
accretion might bring in HVCs with low metallicities.  For Chain A, 
current information is limited to the observed column-density ratios 
$N$(Mg$^+$)/$N$(HI) $>$ 0.035 solar$^{22}$ 
and $N$(Ca$^+$)/$N$(HI) $\sim$ 0.01 solar (see above); in view of possible 
depletion by inclusion into dust grains and uncertain ionization conditions, 
these ratios only give lower limits to total Mg and Ca abundances.  
The best chance for a more significant value lies in measurement of 
the ultraviolet SII lines in Mark 106, since sulphur is not depleted 
onto grains, and S$^+$ is the dominant ionization stage in neutral gas.
Reliable metallicity values and further direct 
distance measurements of HVCs will hold the key to their understanding.  
Since the HVC phenomenon is only loosely defined, and may well include 
objects of very different origins, distance and metallicity measurements of 
various HVCs will be required.


1. Muller C.A., Oort J.H., Raimond E., 
Comptes-rendus Acad. Sci. Paris {\bf 257}, 1661-1664 (1963).

2. Oort J.H., 
Possible interpretations of the high-velocity clouds, 
Bull. Astron. Inst. Netherlands {\bf 18}, 421-438 (1966).

3. Wakker B.P., van Woerden H., 
High-velocity clouds, 
Annual Rev. Astron. Astrophys. {\bf 35}, 217-266 (1997).

4. Bregman J.N., 
The Galactic Fountain of high-velocity clouds, 
Astrophys. J. {\bf 236}, 577-591 (1980).

5. Houck J.C., Bregman J.N., 
Low-temperature galactic fountains, 
Astrophys. J. {\bf 352}, 506-521 (1990).

6. Oort J.H.,
The formation of galaxies and the origin of the high-velocity hydrogen clouds,
Astron. Astrophys. {\bf 7}, 381-404 (1970).

7. Blitz L., Spergel D.N., Teuben P.J., Hartmann L., Burton W.B.,  
High-velocity clouds: Remnants of Local Group formation,  
Bull. Amer. Astron. Soc. {\bf 28}, 1349 (1996).

8. Blitz L., Spergel D.N., Teuben P.J., Hartmann L., Burton W.B.,
High-velocity clouds: Building blocks of the Local Group,
Astrophys. J. {\bf 514}, 818-843 (1999).

9. Kerr F.J., Sullivan W.T.,
The high-velocity hydrogen clouds considered as satellites of the Galaxy,
Astrophys. J. {\bf 158}, 115-122 (1969).

10. Verschuur G.L.,
High-velocity neutral hydrogen,
Annual Rev. Astron. Astrophys. {\bf 13}, 257-293 (1975).

11. Schwarz U.J., Wakker B.P., van Woerden H., 
Distance and metallicity limits of high-velocity clouds,
Astron. Astrophys. {\bf 302}, 364-381 (1995).

12. Danly L., Albert C.E., Kuntz K.D.,
A determination of the distance to the high-velocity cloud Complex M, 
Astrophys. J. {\bf 416}, L29-31 (1993).

13. Ryans R.S.I., Keenan F.P., Sembach K.R., Davies R.D.,
The distance to Complex M and the Intermediate Velocity Arch,
Mon. Not. R. Astron. Soc. {\bf 289}, 83-96 (1997).

14. Bates B., Catney M.G., Keenan F.P.,
High-velocity gas components towards 4 Lac,
Mon. Not. R. Astron. Soc. {\bf 242}, 267-270 (1990).

15. Stoppelenburg P.S., Schwarz U.J., van Woerden H.,
Westerbork HI observations of two high-velocity clouds,
Astron. Astrophys. {\bf 338}, 200-208 (1998).

16. Danly L., Lockman F.J., Meade M.R., Savage B.D.,
Ultraviolet and radio observations of Milky Way halo gas,
Astrophys. J. Suppl. Ser. {\bf 81}, 125-161 (1992).

17. de Boer K.S. \etal,
The distance to the Complex C of high-velocity halo clouds,
Astron. Astrophys. {\bf 286}, 925-934 (1994).

18. Van Woerden H., Peletier R.F., Schwarz U.J., Wakker B.P., Kalberla P.M.W.,
Distances and metallicities of high-velocity clouds,
in Stromlo Workshop on High-Velocity Clouds (eds. Gibson B.K., Putman  M.E.), 
ASP Conf. Ser. {\bf 166}, 1-25 (1999).

19. Tamanaha C.M.,
Distance constraints to the AntiCenter high-velocity clouds,
Astrophys. J. Suppl. Ser. {\bf 104}, 81-100 (1996).

20. Wakker B.P., van Woerden H., de Boer K.S., Kalberla P.M.W., 
A lower limit to the distance of HVC complex H,	
Astrophys. J. {\bf 493}, 762-774 (1998).

21. Wakker B.P., van Woerden H.,
Distribution and origin of high-velocity clouds. III.
	Clouds, complexes and populations,
Astron. Astrophys. {\bf 250}, 509-532 (1991).

22. Wakker B.P., \etal,
The distance to two hydrogen clouds: the high-velocity complex A and the 
	low-latitude Intermediate-Velocity Arch,
Astrophys. J. {\bf 473}, 834-848 (1996).

23. Adelman S.J., Fisher W.A., Hill G., 
An atlas of the field horizontal branch stars HD 64488, HD 109995 
	and HD 161817 in the photographic region,
Publ. Dominion Astrophys. Obs. Victoria {\bf 16}, 203-280 (1987).

24. Wakker B.P., Murphy E.M., van Woerden H., Dame T.M.,
A sensitive search for molecular gas in high-velocity clouds,
Astrophys. J. {\bf 488}, 216-223 (1997).

25. Tufte S.L., Reynolds R.J., Haffner L.M.,
WHAM observations of H$\alpha$ emission from high-velocity clouds in the
M, A, and C complexes,
Astrophys. J. {\bf 504}, 773-784 (1998).

26. Verschuur G.L.,
An association between HI concentrations within high-velocity clouds A and C
	and nearby molecular clouds,
Astrophys. J. {\bf 361}. 497-510 (1990).

27. Meyerdierks H.,
A cloud-Galaxy collision: observation and theory,
Astron. Astrophys. {\bf 251}, 269-275 (1991).

28. Verschuur G.L.,
The high-velocity cloud complexes as extragalactic objects in the Local Group,
Astrophys. J. {\bf 156}, 771-777 (1969).

29. Oort J.H.,
Speculations on the origin of the Chain A of high-velocity clouds,
in "Problems of Physics and the Evolution of the Universe" (ed. Mirzoyan L.),
259-280.
(Yerevan: Academy of Sciences of Armenian SSR, 1978).

30. Wakker B.P., Schwarz U.J.,
Westerbork observations of high-velocity clouds. Discussion,
Astron. Astrophys. {\bf 250}, 484-498 (1991).

31. Benjamin R.A., Danly L.,
High-velocity rain: The terminal velocity model of Galactic infall,
Astrophys. J. {\bf 481}, 764-774 (1997).

32. Adelman S.J.,
Elemental abundance analyses with coadded DAO spectrograms - IV. 
	Revision of previous analyses,
Mon. Not. R. Astron. Soc. {\bf 235}, 749-762 (1988).

33. Chaboyer B., Demarque P., Kernan P.J., Krauss L.M., 
The age of globular clusters in light of Hipparcos: resolving the age problem?, 
Astrophys. J. {\bf 494}, 96-110 (1998).

34. Fernley J., \etal ,
The absolute magnitudes of RR Lyraes from HIPPARCOS parallaxes and proper
	motions,
Astron. Astrophys. {\bf 330}, 515-520 (1998).

35. Lucke P.B.,
The distribution of color excesses and interstellar reddening material in the 
	solar neighborhood,
Astron. Astrophys. {\bf 64}, 367-372 (1978).

36. Hoffmeister C.,
Neuer RR Lyrae-Stern S5218 Ursae Majoris,
Astron. Nachrichten {\bf 284}, 165-166 (1958).

Acknowledgments.
The William Herschel Telescope (WHT) is operated by the Royal Greenwich 
Observatory, in the Observatorio del Roque de los Muchachos of the Instituto 
de Astrof\'\i sica de Canarias, with financial support from PPARC (UK) and
NWO (NL).  We thank the NFRA La Palma Programme Committee, and in 
particular Huib Henrichs, for their support of our program.  The Effelsberg 
Telescope belongs to the Max Planck Institute for Radio Astronomy in Bonn.  
Wakker was partly supported by NASA through STScI, which is operated by 
AURA, Inc.  Wakker also thanks Blair Savage for financial support and 
useful discussions.  We finally thank the referee who drew our attention 
to the spectrum of HD 161817.

Address for correspondence: hugo@astro.rug.nl


\begin{figure}
\mbox{\epsfxsize=17cm \epsfbox{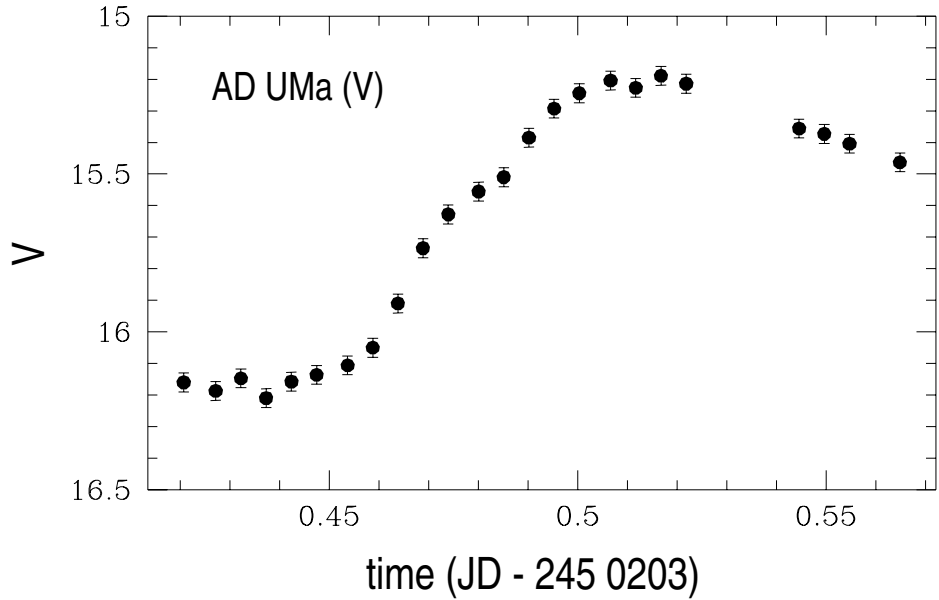}}
\caption{
Partial lightcurve of the RR Lyr star AD UMa, measured in yellow 
light on 1996 Apr 29/30 with the Jacobus Kapteyn Telescope at the Roque 
de los Muchachos Observatory, La Palma TF, Spain by R.F. Peletier, 
H. van Woerden and D. Sprayberry.  Minimum magnitude: 16.17$\pm$0.01, 
maximum: 15.21$\pm$0.01, average 15.69$\pm$0.01 mag.  Assuming 
[Fe/H] = --1.7 by analogy with$^{32}$ HD 161817,
recent calibrations$^{33,34}$ 
of RR Lyr variables based on Hipparcos and other data yield an 
absolute magnitude $M_{\rm V}$ = 0.58$\pm$0.18.  Taking an 
extinction$^{35}$ of 0.1 mag,
the distance of AD UMa implied is $10.1 \pm 0.9$ kpc.  The coordinates
($\alpha$ = $09^{\rm h}23^{\rm m}38.7^{\rm s}$, 
$\delta$ = $+55^{\rm o}46'33''$ (J2000); 
$l$ = 160.40$^{\rm o}$, $b$ = +43.28$^{\rm o}$)
were measured from the Palomar Sky Survey, with reference to the chart 
in the discovery paper$^{36}$;
they differ considerably from those in the Moscow General Catalogue of 
Variable Stars.
}
\end{figure}

\begin{figure}
\mbox{\epsfxsize=17cm  \epsfbox{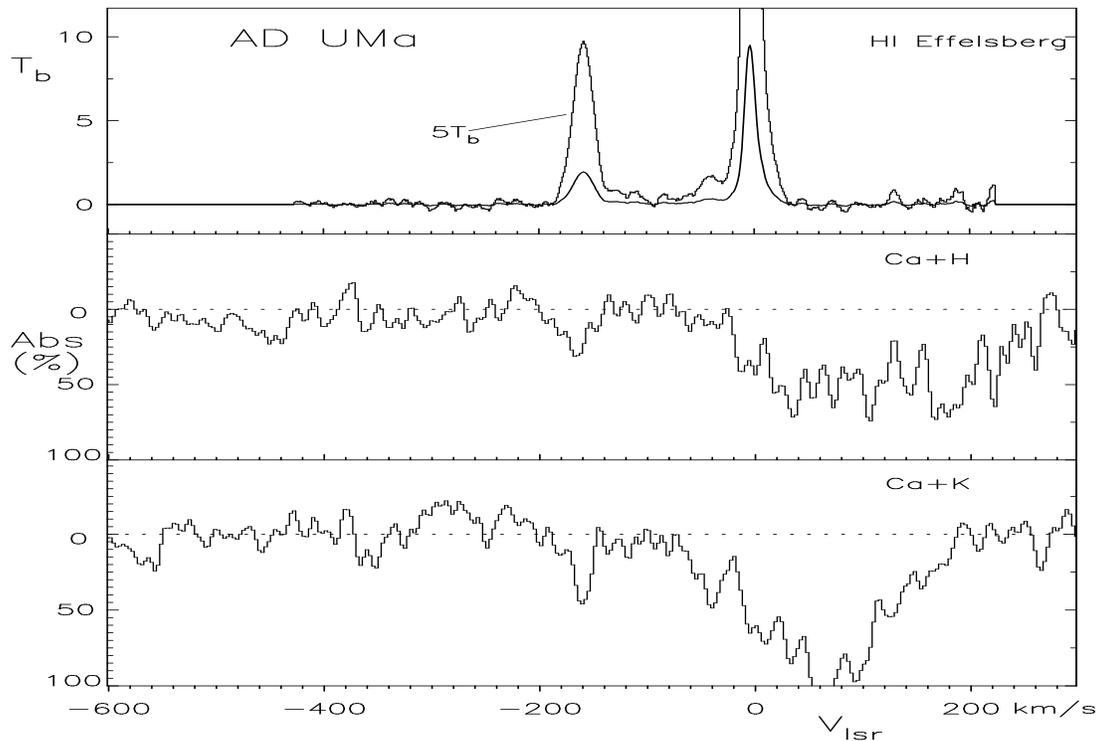}}
\caption{
H{\sc I} emission (top) and Ca{\sc II} absorption measured towards 
the RR Lyr variable AD UMa, which lies at 10 kpc distance, projected on HVC 
Chain A.   The Effelsberg 21-cm profile, measured at 9' beamwidth and 
2 km/s velocity resolution, is shown on two intensity scales, 
different by a factor 5, and has three components, at velocities 
(relative to the local standard of rest, LSR) of - 4, - 40 and -158 km/s; 
the last one is due to the HVC.  The Ca{\sc II} H and K lines, measured 
with the William Herschel Telescope at La Palma at 10 km/s resolution 
during maximum phase on 1997 Jan 18, both show -- in addition to a strong 
stellar component near + 70 km/s, and broad H-$\epsilon$ absorption
longward of the Ca{\sc II} H-line -- the same interstellar components as the 
21-cm line.  This proves that HVC Chain A, as well as the other two clouds, 
lies in front of the star.
}
\end{figure}

\begin{figure}
\mbox{\epsfxsize=17cm  \epsfbox{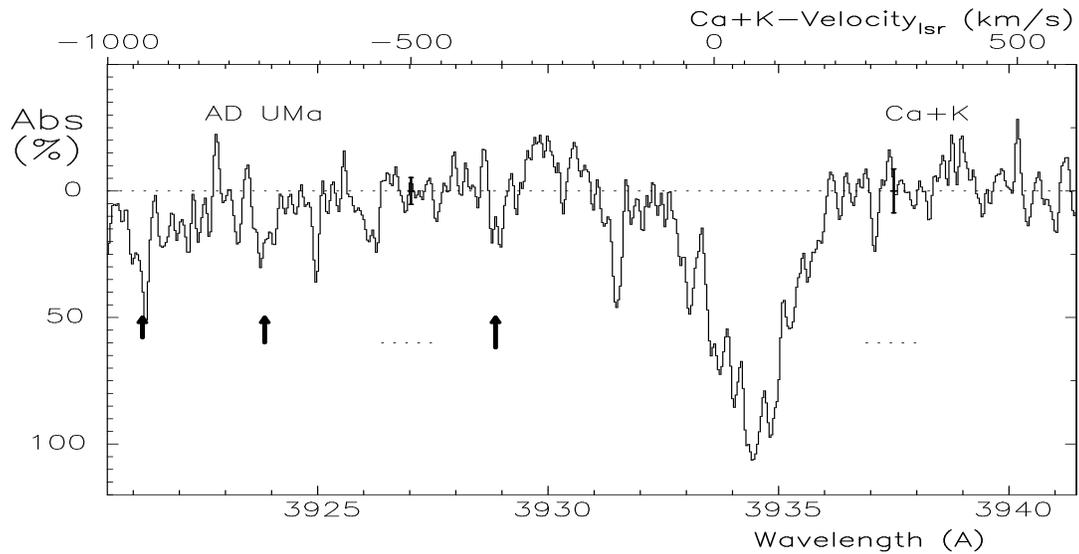}}
\caption{
Spectrum of AD UMa near the Ca{\sc II}-K line 
($\lambda_0$ = 3933.663 $\rm \AA$), taken during maximum phase on 
1997 Jan 18; velocity scale relative to LSR.  Note the stellar K-line 
near + 70 km/s, and the HVC absorption at -- 160 km/s.  Three stellar 
Fe{\sc I} lines, with $\lambda_0$ = 3920.260, 3922.914 and 
3927.922 $\rm \AA$, and shifted by about + 1.0 $\rm \AA$, are indicated by 
arrows.  Gaussian fits yield velocities $v$ = + 76, + 74, and + 81 km/s.  
The average stellar velocity, $+ 77.0 \pm 2.0$ km/s, is consistent with 
values measured for the Fe{\sc I}(4) line at 3860 $\rm \AA$ and, 
more coarsely, for the Ca{\sc II}-K and H-$\delta$ lines.  For the 
Fe{\sc I} line with $\lambda_0$ =  3930.299 $\rm \AA$ see Figure 4.
}
\end{figure}

\begin{figure}
\mbox{\epsfxsize=17cm  \epsfbox{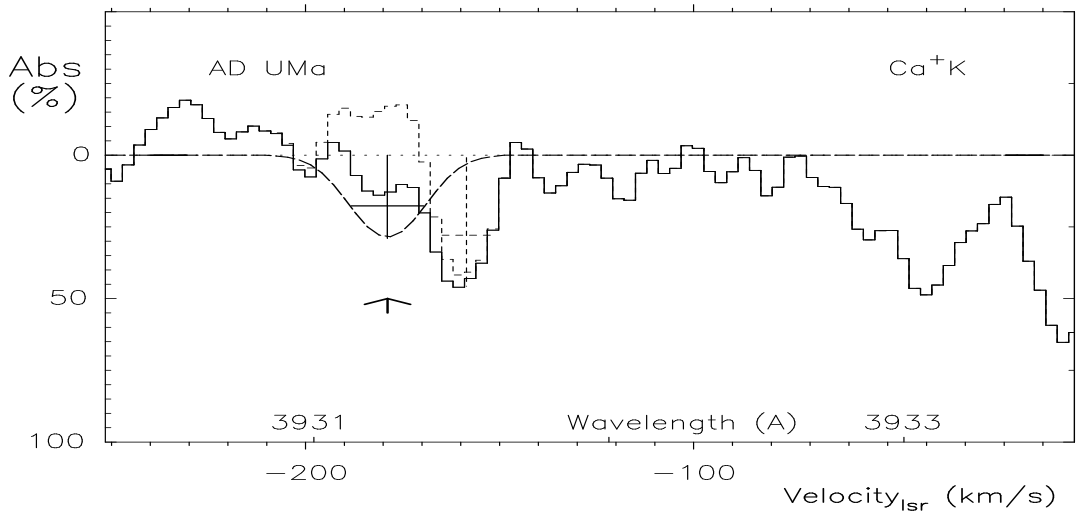}}
\caption{
The predicted profile of the stellar Fe{\sc I} line with 
$\lambda_0$ = 3930.299 $\rm \AA$ is shown by a cross and a dashed 
line.  The observed spectrum (histogram; velocity scale given for 
Ca{\sc II}-K line), in which the continuum is clearly poorly defined, 
barely allows an Fe{\sc I} line of the predicted strength.  Hence,
subtraction of the predicted stellar line from the observed spectrum,
leaving the short-dashed histogram spectrum, sets a lower limit 
to the strength of the neighbouring absorption line.  This deep, narrow 
absorption around velocity -- 158 km/s (relative to LSR), fitted with a 
Gaussian of 6 km/s dispersion (dashed cross), must be due to Ca$^+$-ions 
in HVC Chain A, because it is shifted by $+ 19 \pm 3$ km/s from the 
predicted FeI line. \hfil\break
~~~~~~	In calculating the predicted profile the following data were used:
1) the stellar velocity of + 77 km/s discussed in the caption of Figure 3; 
2) the average velocity dispersion $\sigma$ = 9.3 km/s, found from the 
three Fe{\sc I}(4) lines with $\lambda_0$ = 3920.260, 3922.914 and 
3927.922 $\rm \AA$ (Figure 3); and 
3) an equivalent width W = 90 m$\rm \AA$, derived from the W values 
measured by us for these three lines, together with the ratio of line 
strengths measured$^{25}$ in HD 161817
for the line with $\lambda_0$ = 3930.299 $\rm \AA$ and the other three lines.
The prediction procedure is supported by the fact that the line strengths 
found from Gaussian fits to the spectrum of AD UMa are very similar to 
those$^{25}$ in the bright field-horizontal-branch star HD 161817.}
\end{figure}

\end{document}